\documentclass[10pt,conference]{IEEEtran}

\IEEEoverridecommandlockouts

\usepackage{cite}
\usepackage{amsmath,amssymb,amsfonts}
\usepackage{algorithmic}
\usepackage{graphicx}
\usepackage[hyphens]{url}
\usepackage{textcomp}
\usepackage{fancyhdr}

\usepackage{comment}
\usepackage{multirow}
\usepackage{colortbl}
\usepackage{soul}
\usepackage[table,xcdraw]{xcolor}
\begin{document}

\title{Systolic Array Data Flows for Efficient Matrix Multiplication in Deep Neural Networks\\
}

\author{\IEEEauthorblockN{Tejas Raja}
\IEEEauthorblockA{\textit{Chelmsford High School} \\
\textit
Chelmsford, MA \\
tejasraja21@gmail.com}
}
\pagestyle{plain}

\maketitle

\begin{abstract}

The paper discusses how Systolic Arrays can improve matrix multiplication for deep neural networks (DNNs). With AI models like OpenAI's GPT now containing trillions of parameters, the need for efficient matrix multiplication is more critical than ever. In this paper, the three main systolic array data flows: Weight Stationary (WS), Input Stationary (IS), and Output Stationary (OS) are discussed. Each data flow's energy consumption and efficiency across various matrix sizes are calculated using the SCALE-Sim simulator. The results show that selecting the right data flow for specific matrix configurations can drastically reduce energy consumption. The conclusions provide helpful insights into optimizing hardware for AI and machine learning applications, offering potential improvements in designing energy-efficient DNN accelerators.

\end{abstract}


\section{Introduction}

\begin{figure}[htp]
    \centering
    \includegraphics[width=8cm]{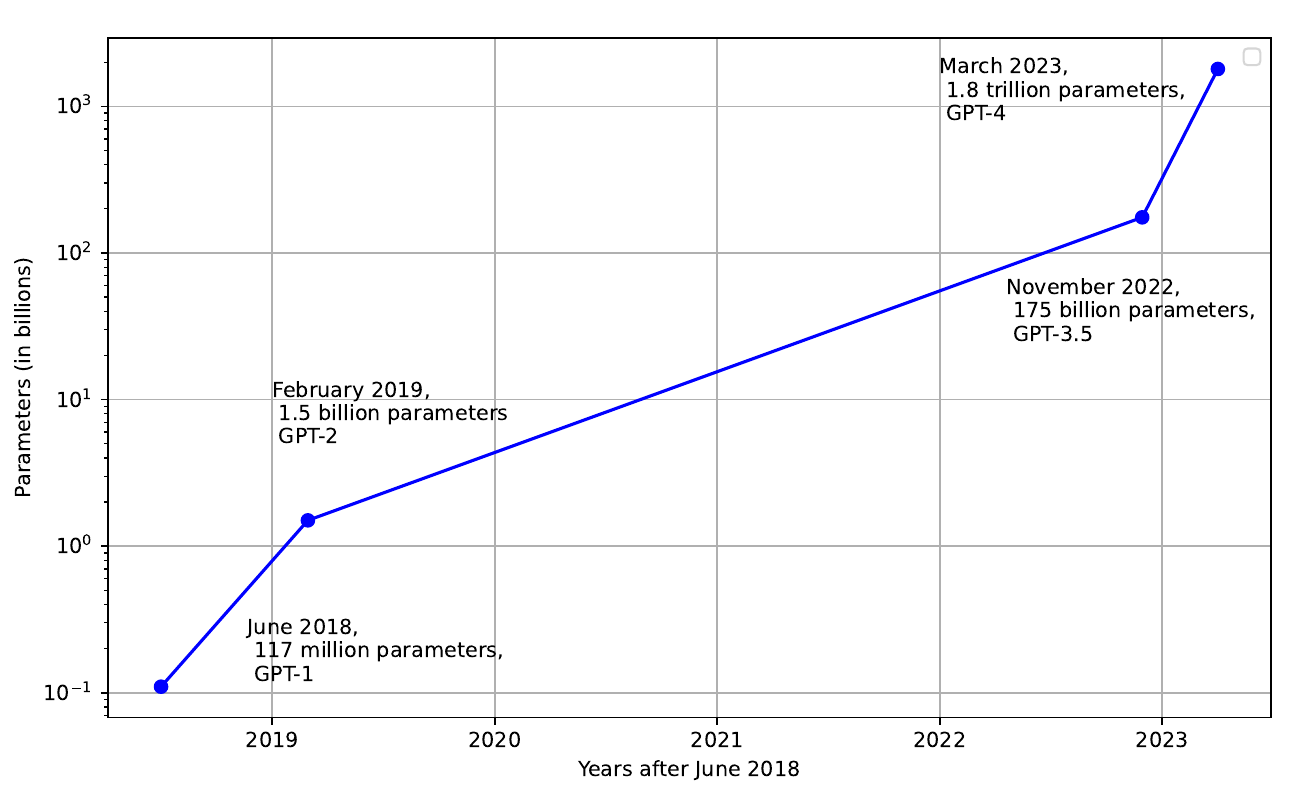}
    \caption{ChatGPT Parameters vs. Year\cite{ChatGPT}}
    \label{fig:ChatGPT_Param}
\end{figure}

In recent years, the increase in the size of artificial intelligence (AI) models has made it critical for large general matrix multiplication (GEMM) operations to be highly energy efficient. For instance, Open AI’s ChatGPT has increased its parameters exponentially over the years \cite{ChatGPT}, starting from 117 million in 2018 to 1.8 trillion parameters in 2023, requiring substantial matrix multiplication during training and inference as shown in Figure \ref{fig:ChatGPT_Param}. Larger matrix sizes require stronger computing hardware. However, the energy and performance costs are also increasing with these large matrix sizes, requiring the design of specified hardware for efficient matrix multiplication.

Traditionally, Von-Neumann architecture-based Central Processing Units (CPU) were used to compute matrix multiplications \cite{arikpo2007neumann}. However, its bottleneck made it challenging to perform at high efficiency. Graphical Processing Units (GPU) were used to combat this, allowing for parallel computation \cite{GPU_Parallel_Computing}. However, as Capra et al. \cite{GPUbadbcvn} point out, GPUs still follow the Von Neumann architecture and therefore suffer from similar memory access bottlenecks, especially in high-demand deep learning applications. Most recently, Google’s Tensor Processing Units (TPU) have emerged as the strongest way to perform extensive matrix operations using systolic arrays that are not limited by Von-Neumann bottlenecks \cite{InDataCenter}. 


Previous works on systolic arrays do not evaluate the specific benefits of using different data flow customizations. The three main data flows are weight stationary (WS), input stationary (IS), and output stationary (OS). Each of the data flows has varied performance and energy efficiency across a range of matrix dimensions and sizes. Having insights into optimal data flow, for given matrix sizes, is crucial in designing and optimizing hardware accelerators for DNN applications.


This paper comprehensively explains the energy required to multiply various matrix sizes with the weight, input, and output stationary data flows. The number of clock cycles required is calculated using the SCALE-Sim \cite{ScaleSim} simulator. The minimum dimension of the systolic array is calculated based on the data flow. The total energy required for each data flow for a range of matrix sizes is calculated. 

\section{Background}
This section discusses the motivation towards specialized hardware for AI/ML computation.
\subsection{Von Neumann Architecture}
\begin{figure}[htp]
    \centering
    \includegraphics[width=8.7cm]{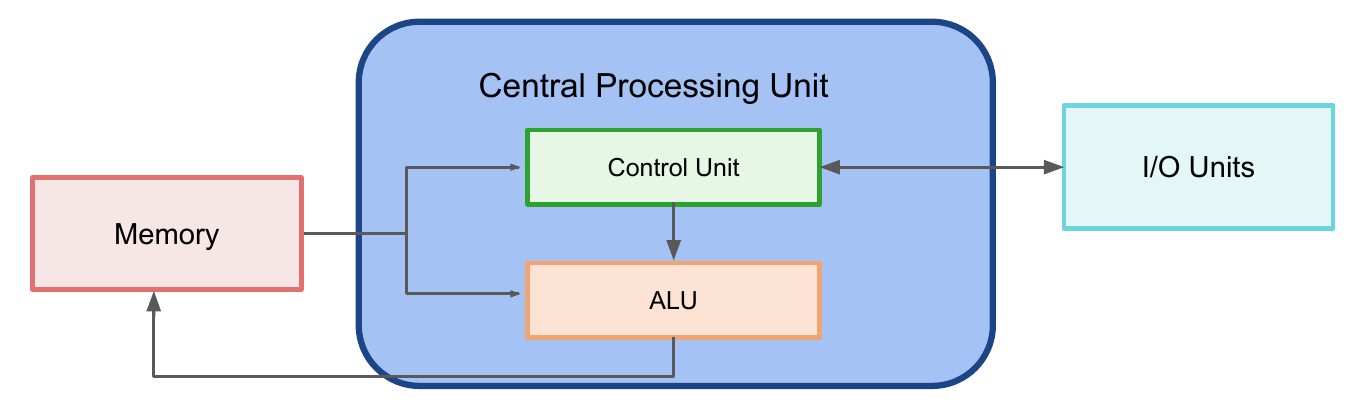}
    \caption{Von Neumann Diagram \cite{arikpo2007neumann}}
    \label{fig:CPUDiagram}
\end{figure}

In 1945, Hungarian mathematician and physicist John Von Neumann created the premise of all computer architecture design: Von Neumann Architecture. Von Neumann architecture executes numerical computation in an independent unit called, CPU, which consists of an arithmetic logic unit (ALU) and a control unit (CU) as shown in Figure \ref{fig:CPUDiagram}. The output and instructions are stored in a separate memory block. When executed, the instructions from memory feed the CPU to perform necessary computations. The output from the CPU is stored back in the central memory unit as shown in Figure \ref{fig:CPUDiagram}. Most computer systems today are based on this architecture and for a good reason. Von Neumann Architecture has a straightforward design, only needing a single memory space for instructions and data. It is also a general-purpose architecture that can handle various applications, from basic computing to executing complex algorithms. Following Moore's Law, such computer architecture’s performance has been scaling until recently. Moore’s law states that the number of transistors in an integrated circuit doubles approximately every two years, ultimately increasing speed while decreasing power consumption and maintaining cost. 

\subsection{Issues With Von Neumann}
Although there are many positives to Von Neumann's architecture, its main setback is its inability to efficiently perform extensive matrix multiplication. Matrix multiplication is crucial today because of its extensive role in scientific computing, computer graphics, data processing, and, most importantly, Deep Neural Networks (DNN) training and inference. Training DNNs involves multiplying a matrix of inputs with a matrix of weights to model the connections between neurons. As DNNs grow, similar to the exponential increase in parameters of large language models (LLM) like ChatGPT, the necessity to perform matrix multiplication efficiently becomes increasingly important.

Von Neumann's architecture is inefficient for matrix multiplication, primarily due to its memory bottleneck. The Von Neumann Bottleneck happens due to the shared bus between the memory and the processing units, forcing the processor to fetch data from the memory sequentially, or one at a a time. This slows down the overall speed of the tasks because the processor is often left to sit idle while the data moves from the memory to the processor. In matrix multiplication, which requires frequent memory access due to the repetitive nature of multiplication and addition, this bottleneck severely limits performance by the back-and-forth data transfer. 

\subsection{CPU vs. GPU vs. TPU}
\begin{figure}[htp]
    \centering
    \includegraphics[width=7.5cm]{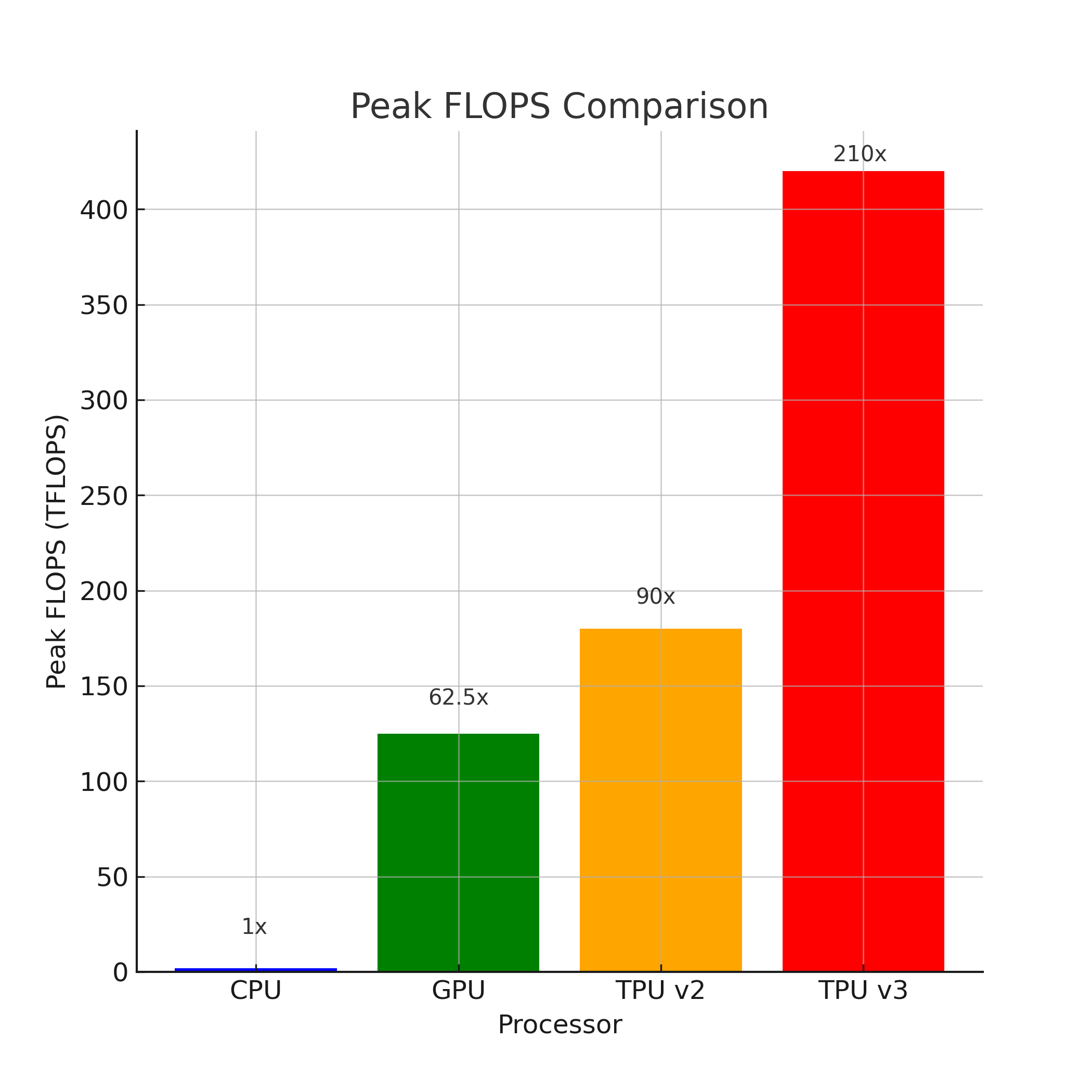}
    \caption{Peak Flops Comparison\cite{DomainSpecific}}
    \label{PeakFlopsCop}
\end{figure}

To tackle the lack of efficiency exhibited by standard Von Neumann designs in performing extensive matrix multiplication operations, an alternative system, called a graphical processing unit (GPU), has been used. A GPU differs from a CPU in that while a CPU is required for all general processing, a GPU is specialized to handle computing and parallel processing \cite{GPU_Parallel_Computing}. Although GPUs are often seen as more efficient for matrix multiplication, they are still built on Von Neumann, leading to the same inefficiencies due to a shared bus and a memory bottleneck. Overcoming these shortcomings, Google's Tensor Processing Unit (TPU) seems to be the most efficient. Figure \ref{PeakFlopsCop} compares the peak floating point operations per second (FLOPS) between each of these processors \cite{DomainSpecific}. Google's TPU differs from standard CPUs and GPUs because of the overall architecture. While CPUs and GPUs use Von Neumann, TPUs use systolic arrays. Unlike traditional Von Neumann architectures, systolic arrays can limit bottlenecks by passing data through a localized array of processing elements (PE) in parallel. 
This allows for Multiplication Accumulate (MAC) operations to be performed simultaneously rather than sequentially, significantly reducing the runtime and energy demands of large matrix multiplications.

\section{Systolic Array Data Flows}

\subsection{Weights, Inputs, and Outputs}

Machine learning (ML) requires the efficient multiplication of two matrices, the input and the weight matrices. In DNN calculations, the inputs are a set of numbers from the activation values of the previous layer or the primary input values. The weights are trainable and data-dependent numbers are used to alter the inputs to predict specific outputs. 

\subsection{What are Systolic Arrays?}
A systolic array is a system made up of a two-dimensional array of units known as processing elements (PE). The operations necessary to multiply matrices are simply specific combinations of multiplication and addition. A processing element has Multiplication and Accumulate (MAC) units for this purpose. The MAC unit multiplies two inputs, corresponding to elements from two matrices, and adds the result to a running total (partial sum). This operation is repeated over time to accumulate results for the output. A processing element also has memory, often noted as local registers, to store incoming data and partial sums. The structure of systolic arrays allows for two major advantages. First, all processing elements can be run simultaneously rather than sequentially. Secondly, the data can be fed in an interleaved way resulting in a pipelined mechanism to compute the output. Systolic arrays are stronger alternatives to traditional Von Neumann as they limit the bottleneck. Systolic arrays do this by not relying on the slow transfer of memory to the processing units to compute tasks sequentially since each processing element in the array has its local memory that can be propagated through each computation. 

\subsection{Mapping Matrices to Systolic Arrays}

\begin{figure}[htp]
    \centering
    \includegraphics[width=6.3cm, height=6.3cm]{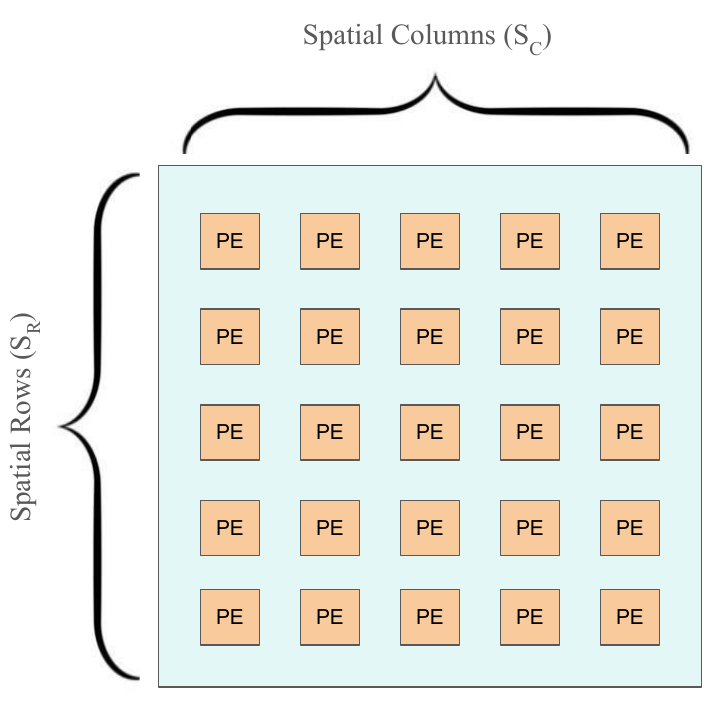}
    \caption{Spatial Rows and Columns to Systolic Array}
    \label{fig:SRSCDiagram}
\end{figure}

To compute matrix multiplications on a systolic array, the matrices are mapped onto the array's physical structure using three dimensions: Spatial Columns ($S_C$), Spatial Rows ($S_R$), and Temporal ($T$). $S_C$ represents the number of columns of processing elements in the array, and $S_R$ represents the number of rows; the spatial dimension corresponds to the physical requirements of the systolic array, as shown in Figure \ref{fig:SRSCDiagram}. $T$, however, refers to the dimension where data from the matrices propagates through the systolic array over time, synchronized by each clock cycle.

\subsection{Data Flows}
\begin{figure}[htp]
    \centering
    \includegraphics[width=8cm]{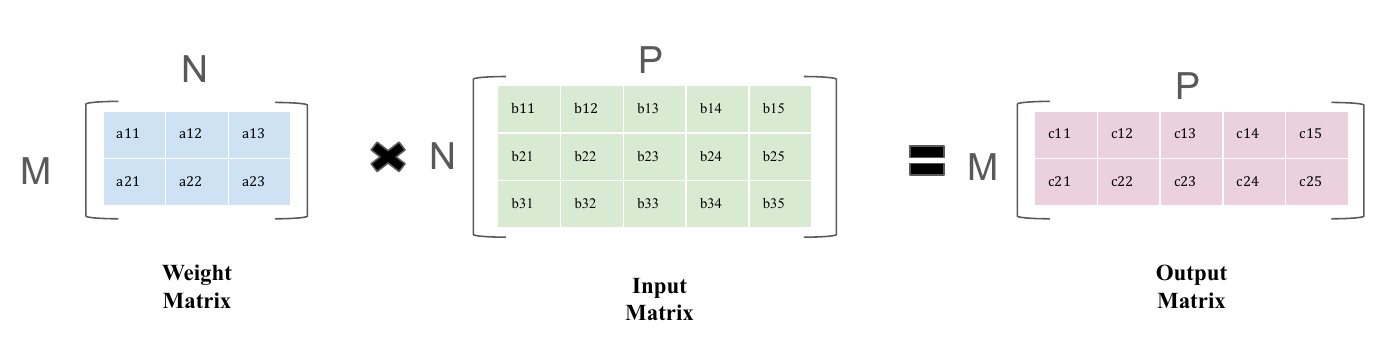}
    \caption{Dimensions of Weight, Input, and Output Matrix}
    \label{fig:MatrixDimensions} 
\end{figure}

Different data flows, such as Weight Stationary (WS), Input Stationary (IS), and Output Stationary (OS), can be used in specific scenarios to optimize the speed and efficiency of the calculations. 
Let the weight matrix be $\mathbf{W} \in \mathbb{R}^{M \times N}$, the input matrix be $\mathbf{I} \in \mathbb{R}^{N \times P}$, and the output matrix be $\mathbf{O} \in \mathbb{R}^{M \times P}$ as shown in Figure \ref{fig:MatrixDimensions}.

\textbf{Weight Stationary}:
In weight stationary (WS) data flow, weight matrix values are pre-filled into the systolic array, while the inputs and partial sums are propagated through the systolic array for each clock cycle. The spatial requirements, or the minimum size of the systolic array, $N_{PE}$, needed is the size of the weight matrix, $M\times N$, while the temporal dimension is $P$.

\textbf{Input Stationary}:
Similar to WS data flow, the IS data flow pre-fills the systolic array with the input matrix values, \(N\times P\). The spatial requirement, $N_{PE}$, for an IS data flow is therefore the size of the input matrix, $N\times P$, while the temporal dimension is mapped to $M$. 

\textbf{Output Stationary}:
Finally, the output stationary data flow is where the output matrix is stationary while both the weight and input values are streamed into the array. In this flow, each PE accumulates the partial sum without propagating it until the entire operation is completed. After the multiplication, the output matrix is read to the output buffer. The spatial dimensions, $N_{PE}$, required to compute matrix multiplication through an OS data flow is the size of the output matrix, $M\times P$, while the temporal dimension is mapped to $N$. 

\subsection{Characterizing Systolic Arrays}

To qualify the data flow for a range of matrix sizes, one needs to estimate the throughput and energy consumption. The total energy consumption is given by Equation \eqref{EnergyEqn}.

\begin{equation}\label{EnergyEqn}
E=N_{PE}\cdot P_{PE}\cdot N_C\cdot T_{clk}
\end{equation}

where $E$ is the total energy, $N_{PE}$ is the number of PEs in the array, $P_{PE}$ is the power consumption per PE, $N_C$ is the total number of cycles needed to compute the matrix multiplication and $T_{clk}$ is the clock period which is $1/F_{clk}$, where $F_{clk}$ is the clock frequency to which the PEs are designed. Samajdar et al, in \cite{ScaleSim}, proposed SCALE-Sim, a cycle simulator to estimate systolic array size and throughput for various data flows. The equation for the number of cycles for input, output, and weight stationary is given by Equation \eqref{TotalCycle}.

\begin{equation}\label{TotalCycle}
N_C=2\cdot S_R+S_C+T-2
\end{equation}

where $S_R$ corresponds to the spatial rows, $S_C$ corresponds to the spatial columns, and $T$ corresponds to the temporal dimension. The total number of PEs is calculated as follows.

\begin{equation}\label{NumPE}
N_{PE}=S_R\cdot S_C
\end{equation}

Based on the type of data flow, the mapping of $S_R$, $S_C$, and $T$ to the matrix dimension are different, as shown in table \ref{tab:maptable}. 
\begin{table}[]
\caption{Mapping of Matrix Dimensions for Systolic Array Data-flows}
\label{tab:maptable}
\begin{tabular}{cccc}
\hline
\rowcolor[HTML]{EFEFEF} 
\textbf{Data flow} & \textbf{\begin{tabular}[c]{@{}c@{}}Spatial \\ Rows ($S_R$)\end{tabular}} & \textbf{\begin{tabular}[c]{@{}c@{}}Spatial \\ Columns ($S_C$)\end{tabular}} & \textbf{\begin{tabular}[c]{@{}c@{}}Temporal \\ ($T$)\end{tabular}} \\ \hline
Weight Stationary (WS) & M & N & P \\ \hline
Input Stationary (IS) & N & P & M \\ \hline
Output Stationary (OS) & M & P & N \\ \hline
\end{tabular}
\end{table}
To determine the most efficient data flow, it is necessary to test different matrix sizes under each data flow.

\section{Evaluation}

In this section, the optimal systolic array data flow for different matrix size configurations is determined by calculating the total energy consumption. For each dimension of the matrices \((M\times N\times P)\), the extreme \(\{min:max\}\) values chosen were \(\{5:500\}\). Based on these extreme values 8 different matrix configurations can be evaluated. 

\begin{figure}[htp]
    \centering
    \includegraphics[width=7.8cm, height=5.4cm]{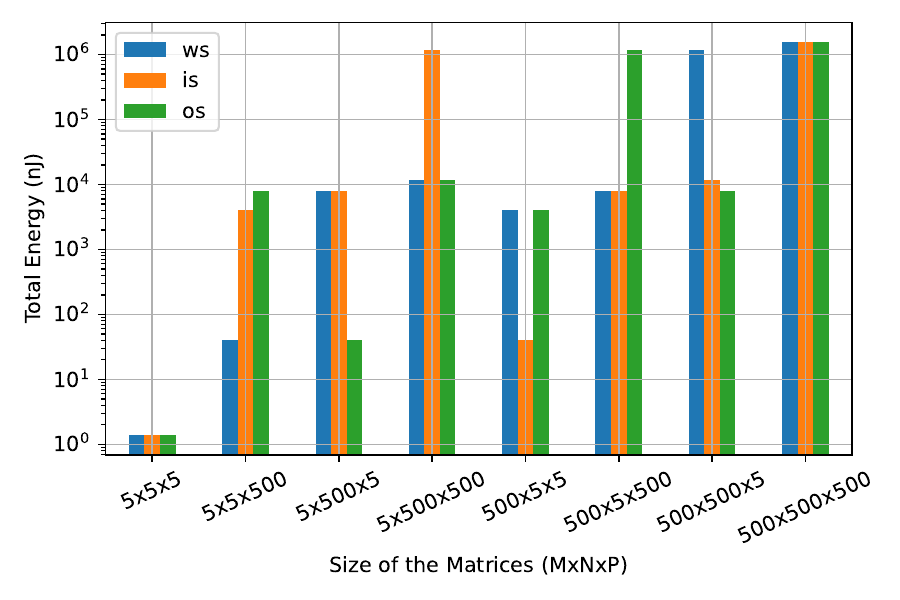}
    \caption{Data Flow Energy Comparison}
    \label{fig:SAnrgcomp}
\end{figure}

A cycle simulator was developed based on the SCALE-Sim paper \cite{ScaleSim} to calculate the total energy required to multiply different configurations of matrix size under each data flow. The findings are plotted in Figure \ref{fig:SAnrgcomp}. The total energy is computed using the Equation \eqref{EnergyEqn}. 
Based on the PE designed by He et. al \cite{SparseTPU} for a 700MHz clock frequency, the power for a regular TPU-based PE, $P_{PE}$, for a 32-bit floating MAC on 28nm technology is 2.17mW. The number of processing elements required is calculated using Equation \eqref{NumPE}, with specific mappings as shown in Table \ref{tab:maptable}. Finally, to find the number of cycles, the Equation \eqref{TotalCycle},  using cycle simulator algorithm \cite{ScaleSim} was used. The total energy is plotted for each configuration of matrix sizes for each of the data flows as shown in figure \ref{fig:SAnrgcomp}.

To reduce the energy required to perform matrix multiplication, the systolic array size needs to be minimal. The number of rows and columns of the systolic array ($S_R, S_C$) that define the array size, as in Equation \eqref{NumPE}, is mapped to the matrix dimensions (\(M, N, P\)) based on the data flow as shown in Table \ref{tab:maptable}. Hence, total energy consumption is minimal when the smallest matrix dimension pair is mapped to the systolic array. The mapping of stationary matrix size to the data flow is defined as $(M\times N)\mapsto WS$, $(N\times P)\mapsto IS$, and $(M\times P)\mapsto OS$.

Based on the graph in Figure \ref{fig:SAnrgcomp}, it is clear that when the three dimensions, (\(M, N, P\)) are equal or similar, all data flows have similar total energy consumption. However, keeping the smaller two dimensions of the matrices (\(M,N,P\)) as the spatial dimensions (\(S_R, S_C\)), will result in the least amount of energy consumed. For example, for configuration \(5\times 5\times 500\), (\(M\times N\times P\)), dimensions (\(M,N\)) are smaller than \(P\), making weight stationary the optimal data flow. 


\section{Related Work}

Batten et al. \cite{LLMBatteneval} shows a promising way to optimize systolic data flows by using PyHDL-Eval, an LLM evaluation framework designed to automate and improve hardware design tasks. By using LLMs to generate efficient hardware configurations and design code, PyHDL-Eval could be adapted to optimize data flows such as Weight Stationary, Input Stationary, and Output Stationary. This approach has the potential to streamline the process of configuring systolic arrays for matrix multiplication in deep neural networks, enhancing both computational efficiency and energy consumption across various matrix dimensions. However, it is important to note that besides optimizing systolic array data flow, key design considerations are memory hierarchy, data interconnection, and multi-core capabilities. 

Besides systolic array-based hardware accelerators, Jung et al. \cite{hammerblade} pioneered Hammerblade, an open-source RISC-V many-core architecture. Being able to scale the number of cores in standard RISC-V processors also significantly increases efficiency for DNN calculations. This shows that as important as it is to focus on the specific cores of processing units, critical advancements can be made by focusing on the System-on-Chip (SoC) design itself.

\section{Conclusion}

Understanding the impact of systolic array data flows is necessary to optimize deep neural network (DNN) accelerators for matrix multiplication. The WS, IS, and OS data flow analysis shows that the choice of data flow significantly affects energy consumption based on matrix dimensions. 
Using the SCALE-Sim-based cycle simulator, it is clear that the smallest dimensions should correspond to the spatial dimensions to require the least amount of energy. This reinforces the need to carefully select data flows depending on matrix configurations to achieve optimal performance in DNN accelerators. With the end of Moore's law \cite{EndOfMoore}, the need for fast processors to train DNNs is more critical now than ever. Thus, optimizing current architectures is an important step towards efficient DNN accelerators. Future work can extend this analysis by exploring advanced data flows and hardware configurations to reduce energy costs further while maintaining high computational efficiency. 

\bibliographystyle{IEEEtranS}
\bibliography{reference}

\end{document}